# Gravitational Waves from the Cosmological Quark-Hadron Phase Transition Revisited

Pauline Lerambert-Potin and José Antonio de Freitas Pacheco

Laboratoire Lagrange,
Université Côte d'Azur,
Observatoire de la Côte d'Azur,
Nice CEDEX 4, 06304, France

pauline.lerambert-potin@etu.univ-cotedazur.fr, pacheco@oca.eu

August 2021

**ABSTRACT**

The recent claim by the NANOGrav collaboration of a possible detection of an isotropic gravitational wave background stimulated a series of investigations searching for the origin of such a signal. The QCD phase transition appears as a natural candidate and in this paper the gravitational spectrum generated during the conversion of quarks into hadrons is calculated. Here, contrary to recent studies, equations of state for the quark-gluon plasma issued from the lattice approach were adopted. The duration of the transition, an important parameter affecting the amplitude of the gravitational wave spectrum, was estimated self-consistently with the dynamics of the universe controlled by the Einstein equations. The gravitational signal generated during the transition peaks around 0.28 µHz with amplitude of $h^2_0 \Omega_{gw} \approx 7.6 \times 10^{-11}$, being unable to explain the claimed NANOGrav signal. However, the expected QCD gravitational wave background could be detected by the planned spatial interferometer Big Bang Observer in its advanced version for frequencies above 1.0 mHz. This possible detection assumes that algorithms recently proposed will be able to disentangle the cosmological signal from that expected for the astrophysical background generated by black hole binaries

## 1 Introduction

The last scattering surface situated around $z_{crit} \sim 1100$ represents a boundary beyond which the universe is opaque to the electromagnetic radiation. In other words, physical processes occurring at redshift higher than $z_{crit}$ cannot be probed by using photons as messengers. However, current fundamental physical theories predict a series of important processes that should have occurred in the primitive universe, like the electroweak or the quark-hadron phase transitions and a putative inflation period necessary to explain, for instance, the observed homogeneity of the cosmic microwave background (CMB).

If the early universe is opaque to the electromagnetic radiation, how could the aforementioned process be probed? Fortunately, the investigation of those events is possible because gravitational waves are simultaneously generated, producing a stochastic cosmological background characterized by a specific spectrum, the smoking-gun of each considered process. These waves interact weakly with matter and, consequently, contrary to photons, may reach present observers with a strong redshifted spectrum.

Gravitational waves were first detected on 14 September 2015 by the laser interferometers LIGO [1], representing a breakthrough on the experimental basis of General Relativity. The signal was produced by the merger of two massive stellar black holes and subsequent detections indicate fusion of binaries involving a pair of black holes (the majority of the cases), a pair of neutron stars [2] or even pairs constituted by a black hole or a neutron star, cases of the sources GW200105 and GW200115. The gravitational wave emission of all possible astrophysical sources along the history of the universe produces also a stochastic background [3], which despite its own interest will interfere with the detection of the cosmological signal that is our main concern in the present study. The LIGO-VIRGO collaboration estimated that at the frequency of about 25 Hz the astrophysical background due to binary coalescences has amplitude of approximately $h^2_0 \Omega_{gw} \approx 10^{-9}$ [4]. However, the spectra of different processes that probably have occurred during the quark-hadron phase transition, generating a cosmological stochastic gravitational wave background, peak at much lower frequencies and their detection requires space based detectors [5]. Recently, the NANOGrav collaboration has claimed the detection of a possible stochastic gravitational wave background [6], although the authors have emphasized the fact that no evidence for quadrupolar spatial correlations have been seen, a property expected to be present in order the signal be consistent with general relativity. The origin of this possible signal is not yet clear but different scenarios have been envisaged in the literature and, in particular, the generation of gravitational waves during the confinement of quarks.

Quantum Chromodynamics (QCD) theory predicts that at high densities and temperatures quarks and gluons are in a state of asymptotic freedom, constituting with leptons and photons the primitive plasma. When the plasma temperature drops below a certain critical value $T_d$, consequence of the expansion of the universe, quarks become confined into hadrons. This process is usually described as a phase transition whose order is still under debate but recent results suggest that the transition could be of first order [7,8]. The physical parameters characterizing the transition depend on the adopted equation of state (EoS), which affects also the properties of the gravitational waves generated during the process.

Recently, it was suggested by [9] that the NANOGrav signal could be explained in terms of a stochastic gravitational wave background generated during the QCD transition. As we will discuss later, the computed signal has some inconsistencies, which have motivated a reanalysis of the expected background generated during the confinement of quarks. In the present work, the gravitational wave spectrum generated during the QCD transition will be computed for two different equations of state issued from the lattice approach. Since the so-called "bag-model" is often referred in the literature, the gravitational signal resulting from this model will be also evaluated for comparison. An improved analysis of the duration of the transition, consistent with the dynamics of the universe described by the Einstein equations will be also presented, since the expected amplitude of the resulting gravitational wave spectrum depends on this parameter. As we shall see later, the predicted gravitational signal issued from the lattice method is unable to explain the NANOGrav signal but this is not the case if the equation of state derived phenomenological bag-model is used. The paper is organized as follows: in Section 2 the various equations of state and the corresponding parameters of the QCD transition will be examined; in Section 3 a self-consistent method to estimate the duration of the transition will be presented; in Section 4 the gravitational wave spectrum resulting from each EoS will be discussed and finally, in Section 5 our main conclusions will be summarized.

# 2 The QCD Phase Transition

In general, two distinct aspects of confinement process must be considered – one associated to chiral symmetry restoration that occurs at a temperature $T_\chi$, which is related to the fact that the QCD Lagrangian has an approximate SU(3) chiral symmetry. Such symmetry is broken by the quark mass term and it holds approximately for the three flavors u, d, s but it is not relevant for the heavier quarks. The chiral symmetry breaking is characterized by a nonzero expectation value of the resulting quark-chiral condensate. For temperatures higher than $T_\chi$ the quark condensate vanishes and the symmetry is restored. Equations of state based on chiral models are also associated to non-null quark chemical potentials (see, for instance [10-12] and references therein), incorporating as well an external pion field. The properties of the QCD transition have also been derived from the Hawking-Page phase transition by using the AdS/QCD correspondence and, in particular, the duration of the transition was assumed to be equal to the evaporation timescale of the equivalent five-dimensional AdS black hole [13,14].

Here, only equations of state based on zero quark and hadron chemical potentials will be considered. As anticipated above, for zero chemical potential, the transition will be of first-order if the mass of the strange quark ms is much larger than the **u**, **d** masses but lower than a critical (uncertain) limit. Above such a critical value the transition is crossover. Calculations made by [7,8] using the Wilson formalism indicate that for **m**$_s$ = 400 MeV the transition is still of first order.

Despite of being "unrealistic", the simplest confinement model and often referred in the literature is the so-called M.I.T. bag-model, in which hadrons are constituted by quarks inside a "bag" confined by the QCD vacuum [15-17]. Inside the bag quarks are in neutral-color bound states and interact among themselves by exchange of an octet of massless colored gluons. The fields describing quarks and gluons do not permeate all space but are rather confined inside the "bag". In the hadron states, quarks occupy the lowest mode of the free Dirac field $\psi$ inside the "bag", supposed to be spherical with radius R [17]. For massless quarks, the resulting pressure and energy density of the quark-gluon plasma are given (in atomic units, $\hbar = c = k = 1$) respectively by

$$P = \frac{\pi\, g_{ef}}{90} T^4 - B \qquad (1)$$

and

$$\rho = \frac{\pi^2 g_{ef}}{30} T^4 + B \qquad (2)$$

where $g_{ef}$ is the effective number of degrees of freedom including both quarks and gluons and B is the energy density of the QCD vacuum, the so-called "bag constant".

The bag-model is based on a phenomenological model but an alternative approach to derive a more "realistic" EoS for the quark-gluon plasma is the lattice gauge theory, which emerged as a successful non-perturbative tool to investigate QCD interactions [18]. Space-time discretization in the lattice involves spacing denoted by *a*, corresponding to a volume $\mathbf{V} = \mathbf{a^3 N^3}$, where **N** is the number of lattice sites along the spatial direction. Reliable predictions from QCD lattice require that the spacing *a* should be small when compared to the QCD scale ($a \approx 0.3$ fm). Since time is imaginary, the temperature is discretized as T = 1/($N_\tau a$), where $N_\tau$ is the number of sites along the time direction with $N^3 N_\tau$ being the lattice hyper-volume [19]. Both the gluon action and the fermion determinant are discretized on the lattice. From the choice of a convenient gauge and fermion operators on the lattice, observable parameters can be derived by using Monte-Carlo techniques. An important result derived from lattice computations is that, indeed, the phase transition is of first order. This is the case if the **u**, **d** quarks have small masses and the strange quark has a mass below a certain critical value [20] as mentioned above. There is a rich literature about equations of state for the quark-gluon plasma derived from the lattice approach (see, for instance, [21-25]. In general, when using the lattice method, the main computed thermodynamic parameter is the so-called trace anomaly I(T), which is the trace of the energy momentum tensor, corresponding to the quantity (ρ − 3P). Notice that either for a pure relativistic free fermion or free boson gases this quantity is zero. The trace anomaly satisfies the equation

$$I(T) = T^5 \frac{d}{dT}\left(\frac{P}{T^4}\right) \quad (3)$$

Therefore, if the trace anomaly as a function of the temperature is known, the pressure can be obtained by integration of the equation above. The quantity **I(T)** has a maximum at a temperature **$T_c$** that satisfies approximately the relation **$T_c \approx (1.1 - 1.2)T_d$**. This maximum, according to different calculations is around **$T_c$** ~200 MeV, indicating that one should expect the confinement at temperatures around 160–180 MeV. Besides the (unrealistic) bag-model, which has the advantage of providing an analytical EoS, two equations of state resulting from lattice computations will be here considered. The first (hereafter lat-1) was computed for temperatures covering the expected transition region, considering physical quark masses (two light and one massive strange quark) and corrected for effects of the chiral condensate [26]. The second EoS (lat-2) includes also three quarks but the authors [27] used an action that minimizes discretization effects on the lattice, computing as well as the correction factors for non-zero chemical potential cases. Here, the adopted values for lat-2 correspond to zero chemical potential. Figure 1 compares these three different equations of state, all of them including three quarks. The EoS for the bag- model was computed for **B** = 58.3 MeV·fm$^{-3}$. Simple inspection of Figure 1 indicates that the bag-model overestimates the pressure for all considered temperatures and tends rapidly to the Boltzmann limit ($P/T^4 \to 5.209$), while such a bound is not attained for the other equations of state even at temperatures around 300 MeV, a value well above the transition temperature as we shall see later. The bag-model is usually computed for massless quarks, one of the reasons explaining the higher pressures.

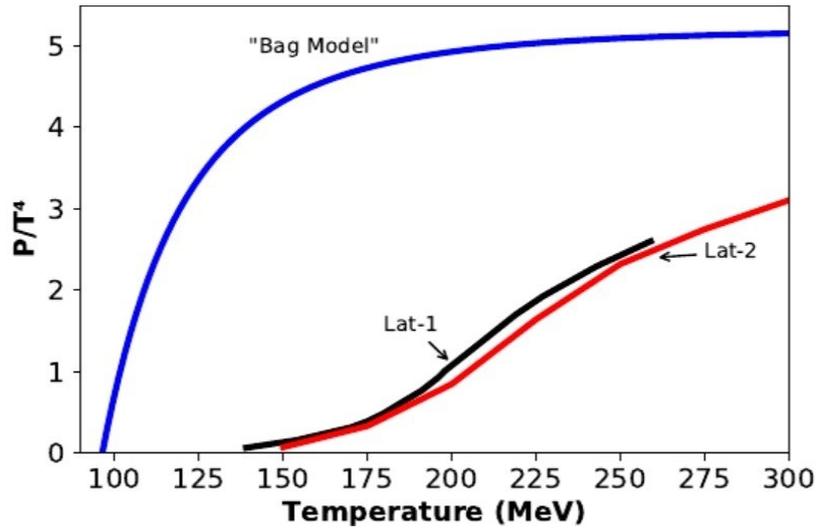

**Figure 1** – Pressure vs. temperature for the three considered EoS mentioned in the text. The bag-model represented by the blue curve was computed using a vacuum energy density of 58.3 MeV·fm$^{-3}$; lat-1 and lat-2 correspond respectively to the black and red curves.

The hadronic phase is generally described by the so-called "Hadron Resonance Gas" (HRG) model [28]. At low energies the HRG is dominated by pions, an approximation that will be adopted in this work. Since s-quarks contribute to the considered quark-gluon plasma, mesons π and K were included in our approach. In this case, the EoS for these mesons (supposed to constitute an ideal bosonic gas) is

$$P_{\pi K} = \frac{T^4}{2\pi^2} \sum_i g_i G_i(T) \tag{4}$$

where the index *i* stands either for pions or kaons, $g_i$ is the number of degrees of freedom of a given particle and the function $G_i(T)$ is given by

$$G_i(T) = -\int_{m_i/T}^{\infty} x^2 \sqrt{\left(x^2 - \frac{m_i^2}{T^2}\right)} \ln(1 - e^{-x}) dx \tag{5}$$

In our calculations, the adopted meson masses were respectively : $\pi^0$ = 134.98 MeV, $\pi^\pm$ = 139.57 MeV, $K^0$ = 497.71 MeV and $K^\pm$ = 493.67 MeV.

The physical parameters of the QCD transition were obtained when the free-energy density F(t) = −P(T) of both phases are equal and they are listed in Table 1. The first column identifies the EoS model, the second and the third columns give the temperature and pressure at the transition while the remaining columns give respectively the energy density of the quark-gluon phase, the energy density of hadrons and the latent heat of the transition, which is simply given by the energy difference between both phases. Notice that the pressure listed in the third column corresponds to the common value for both phases at the transition but does not correspond to the total pressure of the cosmic plasma, since the contribution of leptons and photons was not taken intoaccount.

| Model | $T_d$ (MeV) | $P_d$ (MeV/fm$^3$) | $\rho_{qg}$ (MeV/fm$^3$) | $\rho_{mesons}$ (MeV/fm$^3$) | L (MeV/fm$^3$) |
|---|---|---|---|---|---|
| Bag | 97.4 | 2.73 | 241.4 | 10.8 | 230.6 |
| Lat-1 | 173.8 | 44.24 | 442.7 | 162.6 | 280.1 |
| Lat-2 | 177.7 | 49.11 | 497.0 | 179.9 | 317.0 |

**Table 1:** Physical parameters of the QCD transition derived from the considered equations of state.

Inspection of Table 1 indicates that the transition temperatures derived from the lattice approach are comparable and consistent with the values expected from the analysis of the maximum of the trace anomaly and with values estimated from heavy-ion collisions. The lower temperature obtained for the bag model is a consequence of the steep gradient (dP/dT) observed in Figure 1, resulting from the fact that quarks rapidly attain the state of asymptotic freedom as the temperature increases above the critical value. The energy density at the transition derived from the bag-model is lower than that obtained from the lattice approach essentially because the strong interactions are better described in the later. Notice also that the energy density of quark matter at the transition point derived from the lattice approach is comparable with the value inside a nucleon, $\rho_N \approx 0.45$ GeV·fm$^{-3}$ [24].

## 3 The Duration of the Transition

The QCD transition begins with the appearance of small bubbles of hadrons inside the cosmic plasma. These bubbles grow and coalesce until all quarks and gluons be confined. However, if the mass of the s-quark is above a certain critical value, the transition is "crossover", i.e., quarks are confined smoothly [20,21]. Such a difference brings about important modifications in the gravitational wave spectrum generated during the transition as we shall see later. When hadronic bubbles begin to nucleate, there is a very short period of supercooling. However, the latent heat released as the small bubbles form and expand reheats the cosmic plasma up to the critical transition temperature $T_d$. This short supercooling phase lasts for about 1% of the total duration of the transition and will be neglected in the present calculations.

Along the confinement process the temperature and the pressure remain constant. However the total energy density varies not only due to the expansion of the universe but also because during the transition a variation of the number of the degrees of freedom occurs, from an initial value of $g_{ef}$ = 61.75 down to $g_{ef}$ = 20.25 at the end of the process. The spectrum of the gravitational waves depends on the duration of the conversion of quarks into hadrons and here a simple approach is considered aiming to estimate self-consistently the timescale $\Delta t$ of such a process. Notice that, in general, in the literature this parameter is taken to be $(H\Delta t) = 0.10$ without any consideration about the dynamics of the universe.

We assume that in a given instant of time the fraction of the cosmic plasma in the hadronic phase is $x(t)$, a quantity that varies between 0 and 1 along the conversion of quarks into hadrons. If $\rho_1$ and $\rho_2$ are respectively the total initial and final energy densities, then the energy density at a given instant is

$$\rho(t) = \rho_1(1-x) + \rho_2 x \qquad (6)$$

Using the relation for the latent heat of the transition, that is $L = \rho_1 - \rho_2$, the precedent equation can be recast as

$$\rho(t) = \rho_1 - L x(t) \qquad (7)$$

From the conservation of the energy-momentum, $T^k_{0;k}$, equation and using a Friedman-Robertson Walker metric, one obtains

$$\frac{d\rho}{dt} + 3H(\rho + P) = 0 \qquad (8)$$

where H is the Hubble-Lemaître parameter and P is the total pressure of the cosmic plasma. On the other hand, the Hubble-Lemaître parameter satisfies the equation

$$H^2 = \frac{8\pi G}{3c^2} \rho \qquad (9)$$

Inserting Equation (9) into Equation (8) and making use of Equation (7), one obtains after some algebra an equation for the evolution of the hadronic fraction, i.e.,

$$\frac{dx}{dt} - \sqrt{\frac{24\pi G \rho_1}{\alpha^2 c^2}} (1-\alpha x)^{1/2} (1-\alpha x + \beta) = 0 \qquad (10)$$

where the dimensionless parameters $\alpha = L/\rho_1$ and $\beta = P/\rho_1$ were introduced. In order to have Equation (10) in a complete dimensionless formulation, it is convenient to define a new variable $\tau$ such as $t = \tau t_*$, where the scaling parameter is defined by

$$t_* = \frac{\alpha c}{\sqrt{24\pi G \rho_1}} \tag{11}$$

Hence, Equation (10) can be recast as

$$\frac{dx}{d\tau} - (1-\alpha x)^{1/2}(1-\alpha x + \beta) = 0 \tag{12}$$

Under the assumption that at the beginning of the transition the dimensionless time is equal to zero, the initial condition to be applied on the solution of Equation (12) is x(0) = 0, which gives the solution

$$x(\tau) = \frac{1}{\alpha} - \frac{\beta}{\alpha}\tan^2\left[\arctan\left(\frac{1}{\sqrt{\beta}}\right) - \frac{\alpha\sqrt{\beta}}{2}\tau\right] \tag{13}$$

If the parameters α and β of the transition are known, the dimensionless duration Δτ can be obtained from the condition x(Δτ) = 1 and the true duration from (Δτ)t$_*$. Figure 2 shows the evolution of the energy density for the considered equations of state using Equation (13). Notice that the time was normalized for each case in terms of the corresponding duration.

The parameters α and β as well as the duration of the transition for each considered EoS are shown in Table 2. The different EoS are indicated in the first column; the parameters α and β were computed using the previous results, including the contribution of leptons (electrons, muons, three neutrinos and their anti-particles) and photons, are given respectively in the second and third columns; the duration of the transition in microseconds is listed in the fourth column. Finally, in the last column is shown the ratio between the scale parameters respectively at the beginning and at the end of the transition.

Notice that parameters resulting from the lattice approach are quite comparable but this is not the case for the bag model, which has a transition duration almost one order of magnitude longer. In fact, the (total) pressure is the determinant factor controlling the duration, explaining the differences among the considered equations of state. This will affect significantly the resulting gravitational wave spectrum as we will shown in the next section.

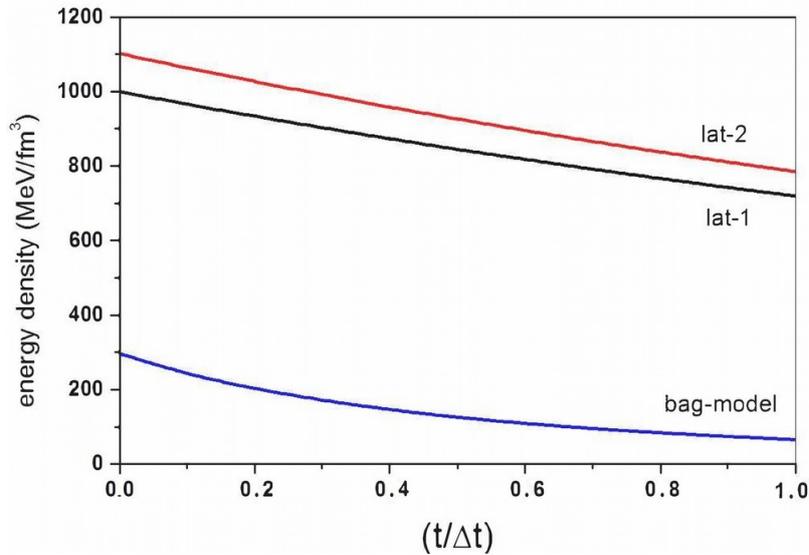

**Figure 2 –** Evolution of the total energy density along the QCD transition for the three considered equations of state. Time was normalized in terms of the corresponding transition duration.

| Model | α | β | Δt (µs) | $a_1/a_2$ |
|---|---|---|---|---|
| Bag | 0.7783 | 0.0710 | 37.11 | 0.6489 |
| Lat-1 | 0.2802 | 0.2299 | 2.97 | 0.9174 |
| Lat-2 | 0.2867 | 0.2279 | 2.91 | 0.9152 |

**Table 2:** Duration and beginning of the QCD transition.

Besides the duration of the transition, it is interesting to compute the variation of the scale factor between the beginning and the end of the process. If the entropy in a unit comoving volume is conserved along the conversion of quarks into hadrons, then the ratio between the initial scale factor $a_1$ and the final value $a_2$ at the end of the transition is given by

$$\frac{a_1}{a_2} = \left(\frac{s_2}{s_1}\right)^{1/3} \quad (14)$$

where s is the entropy density given by s = (ρ + P)/T. Using the definition of the parameters α and β, the equation above can be recast as

$$\frac{a_1}{a_2} = \left(\frac{1+\beta-\alpha}{1+\beta}\right)^{1/3} \quad (15)$$

The resulting values computed from Equation (15) are given in the last column of Table 2.

## 4 Generation of Gravitational Waves

If the transition is not crossover, bubbles constituted by hadrons begin to appear in the cosmic plasma during the confinement process, which grow and coalesce until all quarks (and gluons) be confined. A detailed description of the nucleation and percolation processes can be found, for instance, in [29] and in a more recent study by [30].

Bubbles interact along the transition and are deformed, losing their spherical symmetry and allowing emission of gravitational waves [31,32]. The expansion and the collisions among bubbles generate sound waves and turbulence that are additional sources of gravitational waves. The parameters involved in all these mechanisms depend on the physical conditions of the QCD transition itself. One of these parameters was already defined, corresponds to the ratio between the latent heat of the transition and the total thermal energy, that is, α = L/ρ$_1$. Moreover, the efficiency $\kappa_T$ of the energy transfer from the latent heat to kinetic motions, which depends on the parameter α, can be expressed by a relation based on a fit of numerical calculations [31], i.e.,

$$\kappa_T = \frac{1}{(1+0.715\alpha)}\left[0.715\alpha + \frac{4}{27}\sqrt{\frac{3\alpha}{2}}\right] \quad (16)$$

The wall expansion velocity in the detonation approximation, valid when the bubble walls are thin in comparison with the Hubble radius, was taken from [33]

$$u_\omega = \frac{\sqrt{1/3} + \sqrt{\alpha^2 + 2\alpha/3}}{(1+\alpha)} \tag{17}$$

The expression above also depends essentially on the parameter α and indicates that bubbles expand supersonically ($u_w > \sqrt{1/3}$). Relations given by Equations (16) and (17) are good approximations as far as α < 1.

Generally, in cosmology, the energy density of the different constituents of the Universe are expressed in terms of the so-called critical energy density $\rho_c$, which is defined in terms of the present value of the Hubble-Lemaître parameter, that is

$$H_0^2 = \frac{8\pi G}{3c^2} \rho_c \tag{18}$$

Similarly, the density parameter related to energy density spectrum per logarithm interval of gravitational waves is defined as

$$\Omega_{gw} = \frac{1}{\rho_c} \frac{d\rho_{gw}}{d\ln\nu} \tag{19}$$

For the gravitational wave mechanism involving collision of hadronic bubbles, the spectrum was taken from [34,35], i.e.,

$$h_0^2 \Omega_c(\nu) = 1.67 \times 10^{-5} (H\Delta t)^2 \left(\frac{0.11 u_w^3}{0.42 + u_w^2}\right)\left(\frac{\kappa_T \alpha}{1+\alpha}\right)^2 \left(\frac{100}{g}\right)^{1/3} S_c(\nu) \tag{20}$$

In the equation above, as usual, $h_0$ stands for the Hubble-Lemaître parameter in terms of the "canonical" value of 100 km/s/Mpc and $S_c(\nu)$ is the spectral shape of such a mechanism given by [33,35]

$$s_c(\nu) = 3.8 \left(\frac{\nu}{\nu_c}\right)^{2.8} \left[1 + 2.8\left(\frac{\nu}{\nu_c}\right)^{3.8}\right]^{-1} \tag{21}$$

where the characteristic frequency is defined by

$$\nu_c = 1.64 \times 10^{-8} \left(\frac{0.62}{1.8 - 0.1 u_w + u_w^2}\right)\left(\frac{T_d}{100\,MeV}\right)\left(\frac{g}{100}\right)^{1/6} (H\Delta t)^{-1} Hz \tag{22}$$

When bubbles begin to coalesce, a period of linear fluid evolution appears in which sound waves develop and produce shocks. Such a process is important after the collision phase between bubbles but before the dissipation of the totality of the plasma kinetic energy. The timescale of this process was studied by [36], who estimated that the emission timescale by such a mechanism is of the order of the Hubble timescale. According to [36], the energy transfer efficiency from sound (shock) waves into gravitational wave emission is given by

$$\kappa_{sw} = \left(\frac{\alpha_{eff}}{\alpha}\right)\left(\frac{\alpha_{eff}}{0.73 + \alpha_{eff} + 0.083\sqrt{\alpha_{eff}}}\right) \tag{23}$$

where $\alpha_{eff} = (1 - \kappa_T)\alpha$. The gravitational wave spectrum resulting from this mechanism was taken from [32] and is given by

$$h_0^2 \Omega_{sw} = 2.65 \times 10^{-6} (H \Delta t) \left(\frac{\kappa_{sw} \alpha}{1+\alpha}\right)^2 \left(\frac{100}{g}\right)^{1/3} u_w S_{sw}(\nu) \quad (24)$$

where, the spectral shape is

$$S_{sw}(\nu) = \left(\frac{\nu}{\nu_{sw}}\right)^3 \left[1 + \frac{3}{4}\left(\frac{\nu}{\nu_{sw}}\right)^2\right]^{-7/2} \quad (25)$$

and the characteristic frequency is given by

$$\nu_{sw} = 1.91 \times 10^{-8} \left(\frac{T}{100\,MeV}\right) \left(\frac{g}{100}\right)^{1/6} [u_w (H \Delta t)]^{-1} Hz \quad (26)$$

During the collisional process, the bubbles lose their spherical symmetry and stir the plasma, transferring kinetic energy into eddies having scales comparable to their sizes. If the Reynolds number of the flow is high enough, the kinetic energy injected in eddies cascade into smaller scales, establishing a turbulent spectrum. If the bubble walls are unstable face to the perturbations caused by the turbulent fluid a feedback effect occurs and eddies become an important source of gravitational waves [37,38]. In the present investigation, the adopted gravitational wave spectrum is based on the results of [38]

$$h_0^2 \Omega_t = 3.35 \times 10^{-4} (H \Delta t) \left(\frac{\varepsilon \kappa_T \alpha}{1+\alpha}\right)^{3/2} \left(\frac{100}{g}\right)^{1/3} u_w S_t(\nu) \quad (27)$$

where ε is the fraction of the kinetic energy transferred to turbulent motions that is usually taken equal to 5%. The spectral shape in this case is given by

$$S_t(\nu) = \left(\frac{\nu}{\nu_t}\right)^3 \left\{\left[1 + \left(\frac{\nu}{\nu_t}\right)^{11/3}\right]\left(1 + \frac{\nu}{\nu_*}\right)\right\}^{-1} \quad (28)$$

while the characteristic frequencies in this case are

$$\nu_t = 2.88 \times 10^{-8} \left(\frac{T_d}{100\,MeV}\right) \left(\frac{g}{100\,MeV}\right)^{1/6} [u_w (H \Delta t)]^{-1} Hz \quad (29)$$

and

$$\nu_* = 6.56 \times 10^{-10} \left(\frac{T_d}{100\,MeV}\right) \left(\frac{g}{100}\right)^{1/6} Hz \quad (30)$$

Notice that in these equations, frequencies and amplitudes were already corrected for the redshift. In Equations (24)–(30) appears the parameter g that defines the effective number of degrees of freedom and the Hubble-Lemaître parameter. It is worth mentioning that, since during the transition the temperature remains constant, the total energy density varies essentially as g, which decreases along the confinement process. Consequently, the variation of g can be modelled as that of the energy, i.e., g = $g_1$ −($g_1$ −$g_2$)x(t), with $g_1$ and $g_2$ being respectively the values just before the confinement and just after the end of the process. In this case, a convenient average value can be defined as

$$g_{av} = \frac{1}{\Delta \tau} \int_0^{\Delta t} g(\tau) d\tau \quad (31)$$

The equation above can be evaluated numerically using Equation (13). Similarly, an average for the Hubble-Lemaître parameter can be defined as $H^2_{av} = 8\pi G \rho_{av}/3c^2$, where $\rho_{av}$ is the average energy density defined by an equation similar to Equation (31). In this case, after some trivial algebra, the ratio between the duration of the transition and the Hubble-Lemaître time becomes

$$H_{av} \Delta t = \frac{\alpha}{3}(\Delta \tau)\sqrt{\frac{\rho_{av}}{\rho}} \quad (32)$$

Notice that the right side of Equation (32) is dimensionless as one should expect. Table 3 shows the calculated values of the parameters involved in these equations.

| Model | Bag | Lat-1 | Lat-2 |
|---|---|---|---|
| $\kappa_t$ | 0.4604 | 0.2470 | 0.2507 |
| $\kappa_{sw}$ | 0.1880 | 0.1623 | 0.1637 |
| $\alpha_{eff}$ | 0.4200 | 0.211 | 0.2148 |
| $u_w$ | 0.9210 | 0.8533 | 0.8550 |
| $<g>$ | 34.18 | 39.70 | 39.50 |
| $<H \Delta t>$ | 0.4426 | 0.0863 | 0.0887 |

**Table 3:** Parameters of the gravitational wave spectrum.

Hence, in the present study, having modeled the time evolution of the transition, it was possible to estimate the mean values of g and (HΔt), leading to a more accurate evaluation of the resulting gravitational wave spectrum. In Figure 3 are plotted the gravitational wave background spectra derived from the equations of state discussed previously. For each EoS, the spectrum is the sum of the three considered mechanisms, i.e., collisions, sound waves and turbulence. The expected astrophysical background spectrum due to black hole binaries (BHB) is also displayed. This spectrum was taken from [3] and was corrected for the new values of the local coalescence rate estimated by [1] from Ligo-Virgo data. Also included in the plot the low frequency side of the sensitivity curve of the planned gravitational wave spatial antenna Big Bang Observer (BBO) [40] in its advanced version (BBO-2). The bar labeled NG indicates the possible signal of a gravitational background claimed by the NANO Grav collaboration [6].

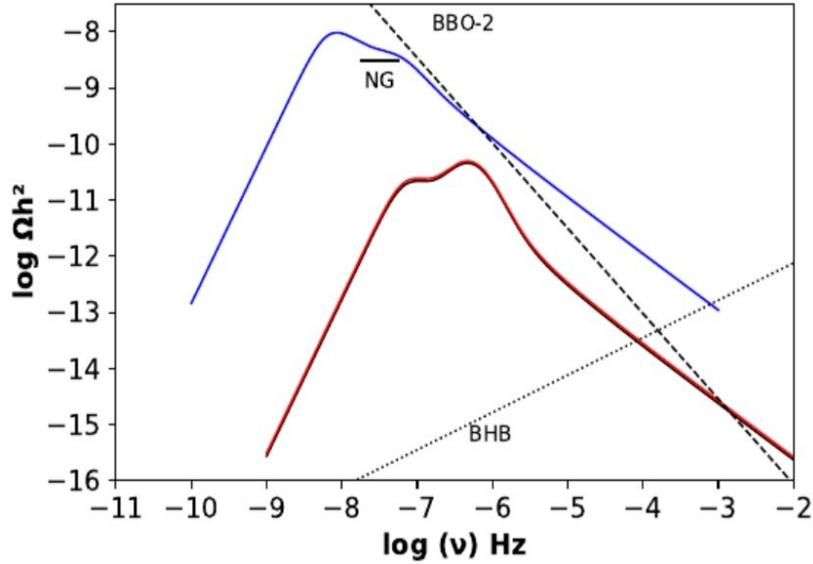

**Figure 3 –** Cosmological gravitational wave background spectra generated during the QCD transition for different equations of state of the quark-gluon plasma: the blue curve corresponds to the bag-model, black and red continuous curves correspond respectively to models lat-1 and lat-2. Also included curves labeled BBO-2 and BHB, which correspond respectively to the sensitivity curve of the planned spatial antenna Big Bang Observer (version two) and to the expected astrophysical background of black holes binaries. The bar labeled NG indicates the possible detection by the NANOGrav experiment.

Inspection of Figure 3 indicates that equations of state issued from the lattice approach lead to quite similar gravitational wave background spectra but differ considerably from that derived from the phenomenological bag-model. The latter has an amplitude almost three orders of magnitude higher and peaks at lower frequencies, since, in this case, the transition occurs at a lower temperature. The large amplitude is due to an important ratio between the latent heat of the transition and the plasma thermal energy, as well as to a longer transition duration in comparison with models issued from the lattice approach. Notice the presence of two "bumps" in the computed spectra shown in Figure 3, which correspond to the contribution of different mechanism: in the bag-model, the first "bump" corresponds to bubble collisions while the second, to sound waves. This is also true for spectra issued from lattice models but the amplitudes are inversed since the transition expected from the bag-model occurs at lower temperature.

Equations of state based on the lattice approach are certainly more "realistic" since they predict transition temperatures consistent with the range of values estimated from heavy-ion collisions, namely, 150–180 MeV, but they are unable to explain the possible gravitational wave background signal reported by the NANOGrav collaboration [6]. On the contrary, the phenomenological bag-model generates a gravitational wave signal consistent with NANOGrav. These two distinct approaches of the QCD thermodynamics can be probed by the planned spatial ex- periment BBO-2. Figure 3 indicates that the gravitational signal produced by the bag-model is above the sensitivity of the BBO-2 antenna for frequencies higher than 580 mHz and above the expected astrophysical background due to black hole binaries. Consequently, this model can be tested in the future. The gravitational signal issued from lattice models is also above the sensitivity of BBO-2 but for frequencies higher than 1.0 mHz. However, in this case, the expected signal is below the predicted astrophysical background, making difficult its detection. Fortunately, new techniques have been recently discussed in the literature, aiming the separation of the cosmological signal from the astrophysical foreground [41-44], opening the possibility to probe not only the QCD transition but also other processes that probably have occurred in the primitive universe.

# 5 Conclusions

In this work the QCD phase transition was revisited using equations of state for deconfined matter based on the lattice approach, including 2 + 1 quarks (two light quarks and one heavy) and zero chemical potential. For comparison reasons, the physical parameters of the QCD transition were also estimated in the case of the phenomenological bag-model, often mentioned in the literature. For all these cases, the cosmological background of gravitational waves generated during the transition was computed.

The duration of the transition Δt is a fundamental parameter defining the amplitude of the gravitational wave spectrum. In the literature, the duration of the transition in terms of the expansion rate of the universe is generally taken equal to (HΔt) = 0.1 (see, for instance, [13,14] and more recently [9]). In the present investigation, by assuming a simple model for the evolution of the total energy during the confinement process, the duration of the transition was estimated self-consistently. Typical (HΔt) values for lattice equations of state are about 0.087, comparable but slightly smaller than the value usually assumed in the literature. Notice that a value approximately five times higher was derived from the bag-model, one of the reasons for the large gravitational wave amplitude derived for this model. Using the time evolution model for the conversion of quarks into hadrons, the average degree of freedom of the particles was estimated, since this is another parameter affecting the amplitude and the characteristic frequencies of the gravitational wave spectrum.

The calculated gravitational wave background issued from the lattice approach is unable to explain the possible NANOGrav signal, confirming the previous but less accurate result by [5]. However, the gravitational wave background derived from the bag-model is able to explain the NANOGrav result since, in this case, the duration of the transition is longer and the ratio between the latent heat and the thermal energy is higher. An encouraging aspect refers to the possibility of a future detection by the planned spatial interferometer BBO-2. The background issued from the bag-model can be detected at frequencies above 580 mHz while the signal resulting from lattice equations of state can be detected above 1.0 mHz. However, in the latter case the signal must be disentangled from the astrophysical background produced by black hole binaries [4] but, as emphasized previously, different algorithms have been proposed to solve this problem. Recently, in [9] the authors considered scenarios in which the deconfined phase is constituted by massive quarks either with zero or nonzero chemical potential. A pure gluon system was also examined as a possible description of the high temperature phase. The equations of state were computed from the approach discussed in [13,14], that is the holographic QCD method. Under these conditions, they have obtained gravitational wave spectra with amplitudes up to $h^2_0\Omega_{gw} \sim 10^{-6.5}$, which are able to explain the NANOGrav signal. This signal is strong enough to be detected by the planned BBO-2 experiment, which will confirm or not these predictions. From their approach [9], the authors estimated transition temperatures in the range 112–192 MeV for quark models and about 255 MeV for a pure gluon description. However the effective number of degrees of freedom and the parameter (HΔt) were fixed a priori and, consequently, these calculations are not self-consistent since the thermodynamic evolution of the transition must be compatible with the dynamics of the universe as discussed in this work. We hope that the planned BBO-2 experiment could give an important contribution by testing these different predictions.

**Acknowledgment** P.L.-P. acknowledges the Laboratory Lagrange for its support during the realization of this work.

**Author Contributions** Conceptualization, P.L.-P. and J.A.d.F.P.; methodology, P.L.-P. and J.A.d.F.P.; software, P.L.-P. and J.A.d.F.P.; validation, P.L.-P. and J.A.d.F.P.; formal analysis, P.L.-P. and J.A.d.F.P.; investigation, P.L.-P. and J.A.d.F.P. resources, P.L.-P. and J.A.d.F.P.; data curation, P.L.-P. and J.A.d.F.P.; writing—original draft preparation, P.L.-P. and J.A.d.F.P.; writing—review and editing, P.L.-P. and J.A.d.F.P.; visualization, P.L.-P. and J.A.d.F.P.; supervision, P.L.-P. and J.A.d.F.P.; project administration, P.L.-P. and J.A.d.F.P.; funding acquisition, P.L.-P. and J.A.d.F.P. All authors have read and agreed to the published version of the manuscript.

**Funding** This research received no external funding.

**Conflicts of Interest** The authors declare no conflict of interest.

# Références